# Seeing lens imaging as a superposition of multiple views

Sascha Grusche

Physikdidaktik, Pädagogische Hochschule Weingarten, Kirchplatz 2, 88250 Weingarten

In the conventional approach to lens imaging, rays are used to map object points to image points. However, many students have a need to think of the image as a whole. To answer this need, lens imaging is reinterpreted as a superposition of sharp images from different viewpoints. These so-called elemental images are uncovered by covering the lens with a pinhole array. Rays are introduced to connect elemental images. Lens ray diagrams are constructed based on bundles of elemental images. The conventional construction method is included as a special case. The proposed approach proceeds from concrete images to abstract rays.

**1. Introduction**

The conventional approach to lens imaging goes back to the German astronomer Johannes Kepler [1,2]. In his view, an extended object consists of several object points, see Fig 1(a). Each object point emits light rays; a corresponding image point is formed where the lens makes these diverging rays converge. Kepler's *point-to-point* approach has been adopted by scientists [3,4], textbook authors [5-10], and teachers [11-17] around the world.

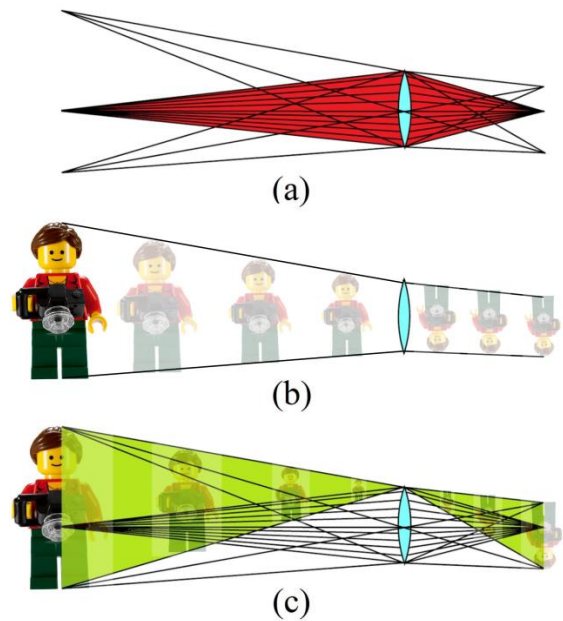

Fig. 1. Three approaches to lens imaging.[1] (a) Conventional point-to-point approach (drawing adapted from Kepler's figure 11 in *Dioptrik* [1]). (b) Students' holistic approach, *cf.* [17]. (Concepts may vary among individuals.) (c) The proposed multi-view approach is based on a reinterpretation of Kepler's ray drawing.

Unfortunately, this point-to-point approach is too abstract for many students. According to empirical studies, many students have a need to think of the image as a whole [11,18-21], see Fig. 1(b). With such a *holistic* approach, many students interpret the rays of geometrical optics as rails that carry the image from the object to the screen [11,19-21].

---

[1] The photo of the toy figure is from *http://lego.wikia.com/wiki/Press_Woman?file=10224fig4.jpg*

Although the students' holistic approach seems naïve, we can find a kernel of truth in it by treating Kepler's ray drawing as an ambiguous image, see Fig. 1(c): Once we switch our attention to rays that go through a single point on the lens, we see that these rays represent a refracted camera obscura projection, *cf.* [22]. Thus, each point on the lens produces a whole image. Images from different points on the lens represent different views [22-24]. This *multi-view* approach allows us to take the students' preconceptions seriously: We may consider rays as connections between camera obscura images, as in Fig. 1(c).

Accordingly, I will use this multi-view approach to build a bridge between the students' holistic approach and the scientists' point-to-point approach. First, I present experiments that allow students to observe the camera obscura images and their superposition. Then, I will simulate lens imaging as a superposition of multiple views. Afterwards, I introduce rays as connections between the camera obscura images. Finally, I propose a method for constructing lens ray diagrams based on these images.

**2. Observing elemental images and their superposition**

Each of our eyes has a lens, so we will start with that.

- Facing a varied background, hold a pen about 30 cm in front of you. Close one eye. With the other eye, try to get a sharp image of the pen and the background simultaneously. It is impossible: If the pen appears sharp, the background looks blurry, and vice versa [25].

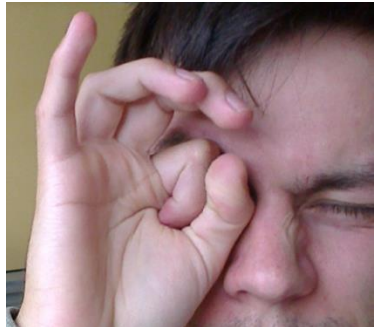

Fig. 2. Fingers forming a pinhole. If the pinhole is moved across the eye, the perspective changes.

- Curling up the thumb and index finger of your other hand, form a pinhole directly in front of your eye, as in Fig. 2. Through this pinhole, the pen and background appear sharp simultaneously, *cf.* [26].
- If you move the pinhole left and right or up and down, the perspective changes as if you move your head in those directions! Does the uncovered eye lens produce multiple views at once?

To answer this question, we build a simple eye model: A convex glass lens represents the eye lens (and other refractive media of the eye), a translucent screen represents the retina, see Fig. 3, *cf.* [25,26]. In front of the eye model, we set up a still life illuminated by white LED lamps. We place an apple so that a sharp image of it appears on the screen; a candle in front of the apple appears blurry on the screen, see Fig. 4(a).

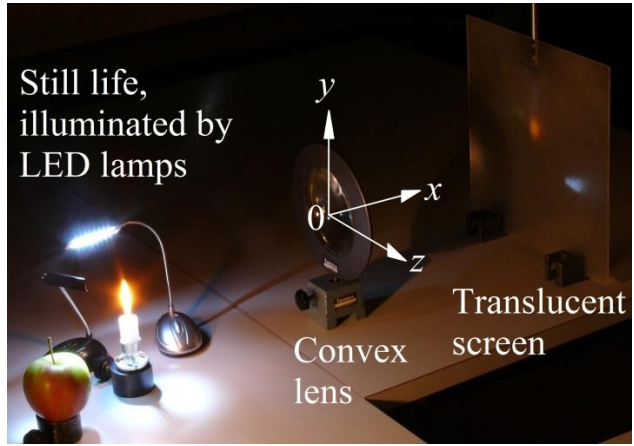

Fig. 3. Simple eye model facing a still life. To avoid stray light, the lens will be surrounded by cardboard. The coordinate system is centered on the lens. Screen distance $x_{screen} = +32$ cm, focal length $f = +20$ cm, $x_{candle} = -28$ cm, $x_{apple} = -40$ cm.

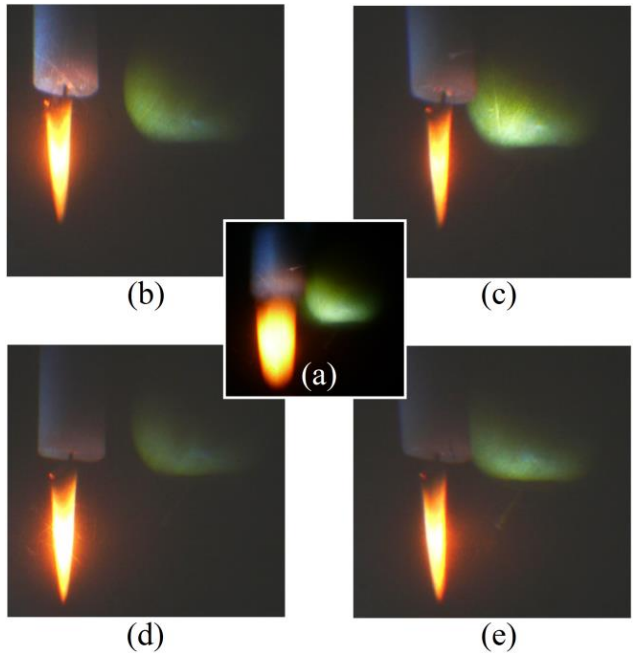

Fig. 4. Moving a pinhole in front of the eye model. (a) Without the pinhole, the apple appears sharp, but the candle appears blurry. (b)-(e) With the pinhole, all objects appear sharp at once, but the perspective changes according to the pinhole position $P_H = (y_H, z_H)$. (b) $P_H = (+2$ cm$, +2$ cm$)$. (c) $P_H = (+2$ cm$, -2$ cm$)$. (d) $P_H = (-2$ cm$, +2$ cm$)$. (e) $P_H = (-2$ cm$, -2$ cm$)$. Pinhole diameter $d = 3$ mm. Photos taken with a Panasonic DMC FZ-50 (aperture number $f/3.6$, exposure time 1/3 s for (a) and 8 s for (b)-(e)).

If we hold a sheet of paper pierced with a pinhole directly in front of the lens, the image of the candle becomes sharp, too, see Fig. 4(b). If we move the pinhole across the lens, the image of the candle moves accordingly—inside the formerly blurry image—, while the image of the apple remains fixed, see Fig. 4(b)-(e). In other words: the perspective changes.

The perspective corresponds to the view from the pinhole: Whatever we can see through the pinhole will appear on the screen behind it. If we replace the moving pinhole with a static pinhole array, as in Fig. 5(a), we get the different perspectives in superposition, see Fig. 5(b). Indeed, the lens produces multiple views at once!

Now, we will do something that the eye cannot do: We will change the distance between the lens and screen. Close behind the lens and pinhole array, the images with different perspective lie side by side, still relatively sharp, see Fig. 5(c). In Integral Imaging [22-24,27,28] (see Section 6), sharp images with different perspective are called *elemental images*, so we will adopt this term.

When we move the projection screen away from the convex lens, the elemental images become larger and ultimately pass across each other, see Fig. 5(d)-(e). (With a concave lens, the elemental images move away from each other.) Accordingly, we may interpret lens imaging as a superposition of elemental images.

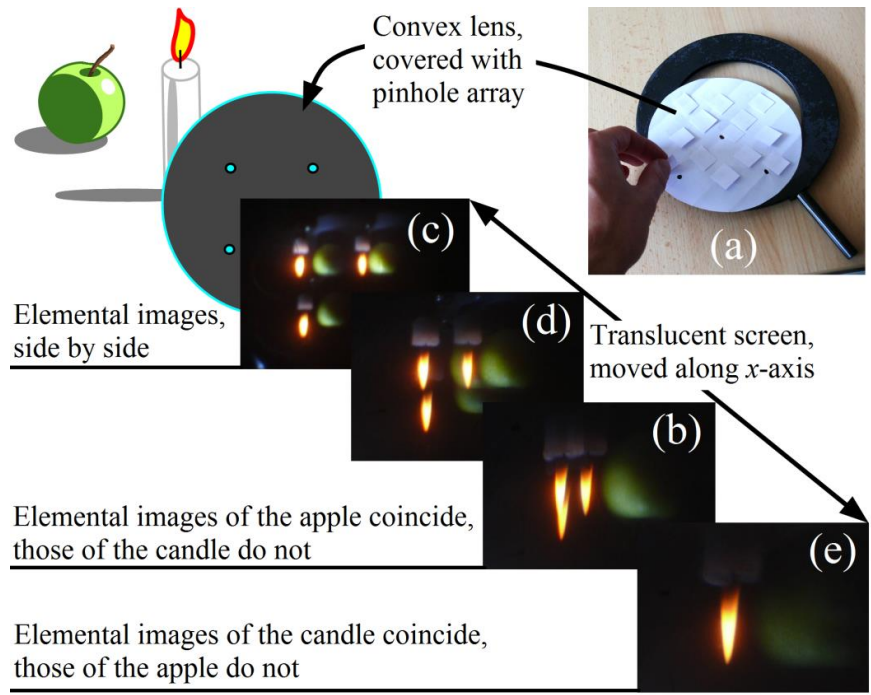

Fig. 5. Observing the superposition of elemental images. (a) To uncover elemental images, the lens is covered with an array of pinholes, which can be individually closed if desired. For (b)-(e), only three pinholes are opened (at ($y$ = +2 cm, $z$ = +2 cm), ($y$ = +2 cm, $z$ = -2 cm), and ($y$ = -2 cm, $z$ = +2 cm), pinhole diameter $d$ = 3 mm). The translucent screen is moved to various screen positions $x_S$. (c) $x_S$ = 8 cm. (d) $x_S$ = 23 cm. (b) $x_S$ = 31 cm. (e) $x_S$ = 42 cm. The photos of the screen were taken with a Panasonic DMC FZ-50 (aperture number $f/3.6$, exposure time 8 s).

The extent to which the elemental images overlap will determine how sharp or blurry the composite image becomes: Where the elemental images are mutually shifted, the composite image is blurry. Only where the elemental images coincide, the composite image is sharp.

Because elemental images from different points on the lens represent different views, it is impossible to bring all of their features into complete overlap at once: For a given lens and given object distance, elemental images coincide only at the so-called image distance. Conversely, at a given screen distance, elemental images coincide only for objects at a specific object distance, depending on the lens. With the uncovered lens, we cannot get a sharp image of the foreground and background simultaneously because their elemental images do not match completely.

## 3. Simulating lens imaging as a superposition of elemental images

Now that we understand lens imaging as a superposition of sharp images with different perspective, we may simulate it accordingly: First, we use a cell phone camera to capture multiple views of a scene, see Fig. 6(a)-(c). Then, we use multiple projectors to superimpose the photos on a projection screen.

If you do not have multiple projectors, you can place multiple mirrors in front of a single projector, each mirror reflecting one of the photos, see Fig. 6(d). With only three photos and three mirrors, the simulation is already realistic: Depending on the screen distance, objects at a certain distance appear sharp in the composite photo, whereas others appear blurry, see Fig. 7.

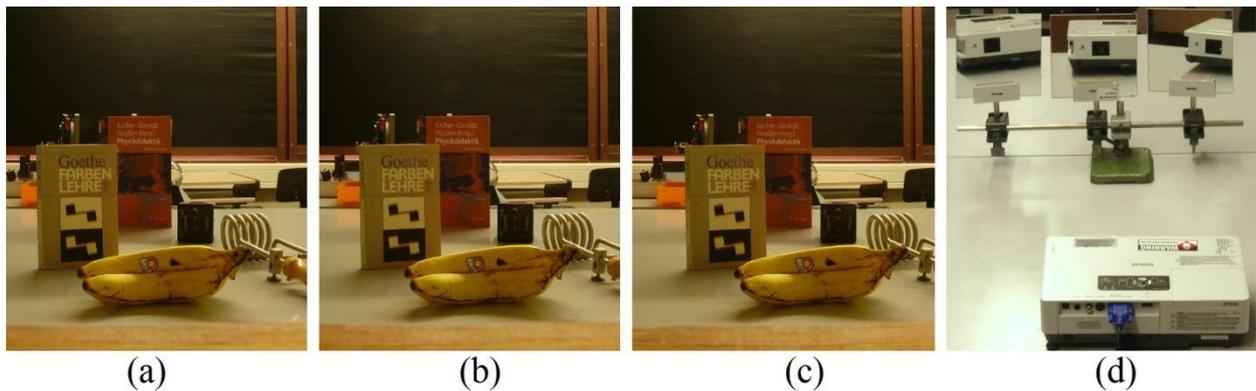

Fig. 6. Photographing and projecting different views. (a)-(c) Photographs of a still life, taken with a cell phone camera at different horizontal positions $z$. (a) $z = -2$ cm. (b) $z = 0$ cm. (c) $z = +2$ cm. (d) Three angled mirrors in front of a single projector are used instead of three angled projectors, as seen from the projection screen.

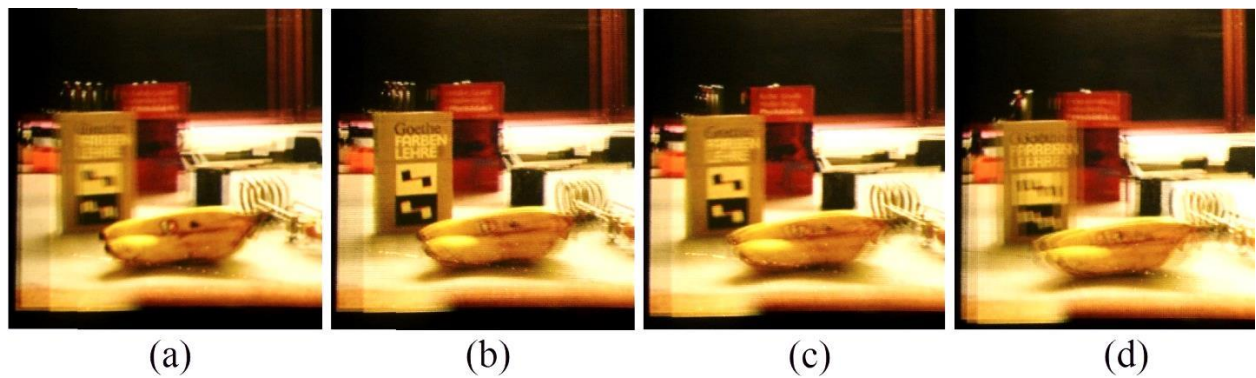

Fig 7. Simulating lens imaging as a superposition of different projections. The photos from Fig. 6(a)-(c) are projected onto a screen via the mirror array in Fig. 6(d). As the screen distance is reduced from (a) to (d), the sharply imaged plane moves from foreground to background: (a) All three projections of the bananas coincide; the projections of other object planes do not. (b) All projections of the gray book coincide. (c) All projections of the orange book coincide. (d) All projections of the background coincide.

Students may craft their own, take-home simulators, see Fig. 8. First, they draw one elemental image onto paper and two other elemental images onto transparencies. (Teachers may help by providing worksheets with photos from different viewpoints.) Then, the students slide the transparencies across the paper to create different conditions of superposition. With this device, students can simulate the effect of a lens with variable optical power, such as the eye lens: If the elemental images of any object are perfectly overlapping, the elemental images of objects at other distances are mutually shifted.

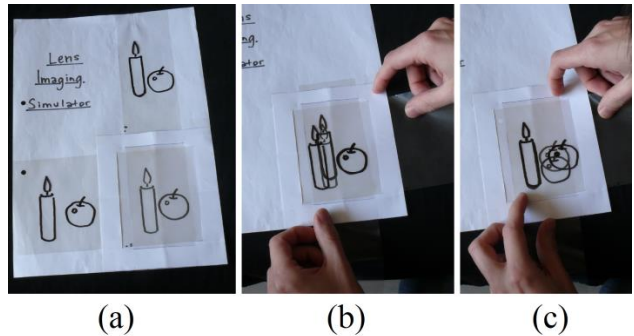

(a)          (b)          (c)

Fig. 8. Take-home simulator. (a) Elemental images with different perspective are displayed. (b) The transparencies are moved over the paper drawing to make the elemental images of the apple overlap. (c) The transparencies are moved further to make the elemental images of the candle overlap.

**4. Using rays to locate elemental images**

For a quantitative treatment of lens imaging, we need to specify the positions of elemental images. To build a bridge to Kepler's ray diagram, we must consider elemental images from points *inside* the lens, *cf.* Figs. 1(a) and (c). Accordingly, we put our pinhole array inside a sandwich of two plano-convex lenses, see Fig. 9(a). For the elemental images to be simple, we place only one object in front of the lens.

To record the positions of elemental images, we trace them on transparencies clipped onto the backside of the translucent screen, see Fig. 9(b). Alternatively, we may paste a transparency with scale markings onto the screen, and simply read off the positions. Based on the measured positions, we transfer the elemental images into a side-view representation, see Fig. 9(c).

In the side-view representation, we note that the size of an elemental image is proportional to its distance from the lens. Hence, we may draw rays between each hole and the corresponding elemental images, see Fig. 9(d). Likewise, we draw rays between each hole and the object, see Fig. 9(d). Do these rays connect elemental images in front of the lens? We hypothesize that they do, *cf.* Fig. 1(c). After all, the plane of a projected, sharp composite image and the corresponding object plane are interchangeable [13]. To verify our hypothesis, we place a pinhole camera before the lens, facing the object: different elemental images contribute different image spots to the pinhole, composing a new image behind the pinhole, *cf.* [22]. In this sense, rays connect elemental images behind and in front of the lens.

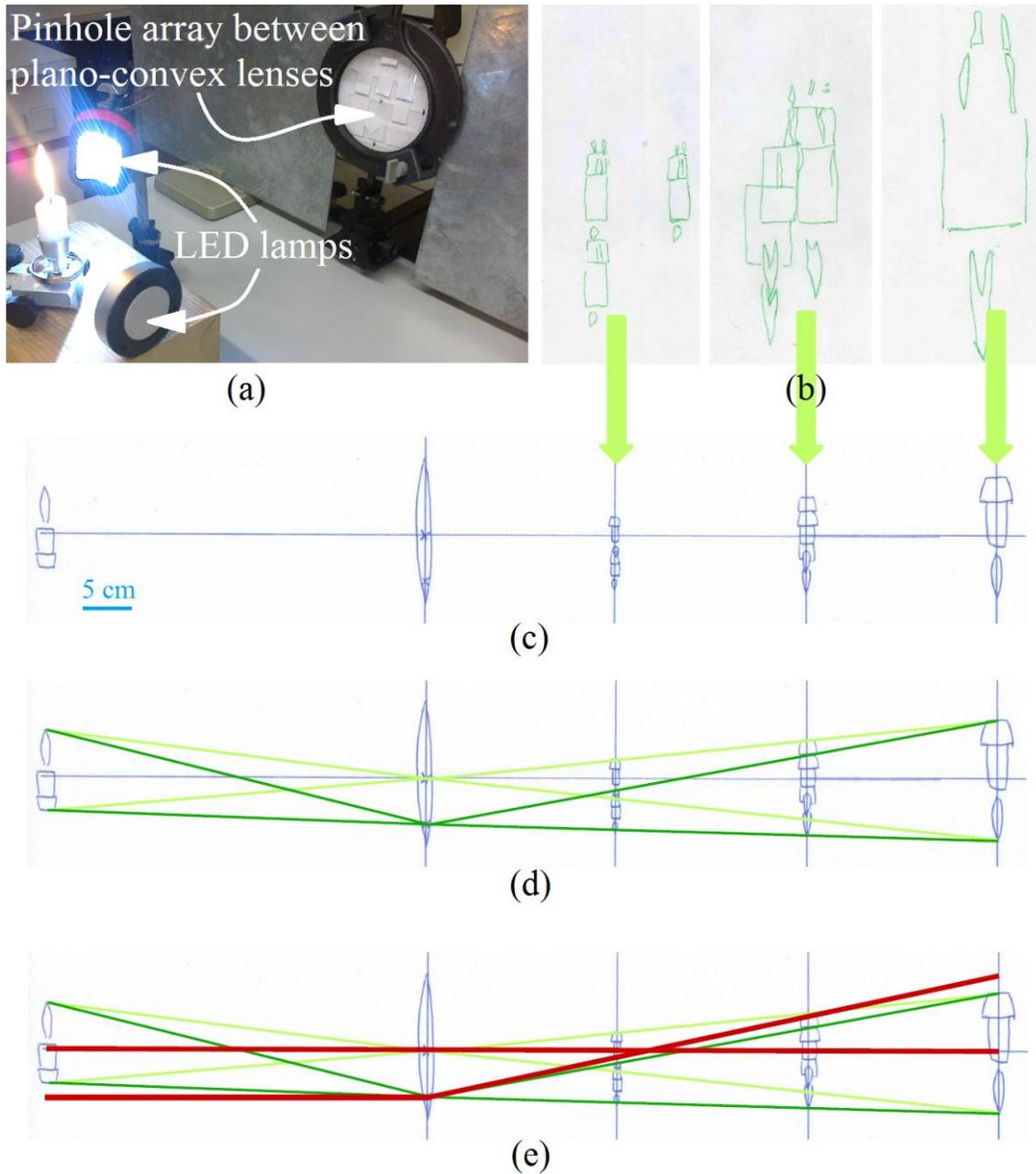

Fig. 9. From images to rays. (a) Experimental setup. Here, each plano-convex lens has a focal length $f =$ +50 cm. (b) On a screen behind the lens, elemental images of the candle are traced on a transparency. (c) The tracings of elemental images are transferred into a side-view representation of the setup. (d) Ray bundles are drawn from each pinhole to the object, and to the corresponding elemental images. (e) From the pinholes, horizontal rays are drawn toward certain object points, and appropriately angled rays are drawn through the corresponding points of the elemental images. The rays intersect in the focal point.

Based on these rays, we come to the following conclusions:
- At a screen distance equal to the object distance, the size of an elemental image is equal to the size of the object.
- An elemental image at a given distance in front of the lens has the same size as an elemental image at the same distance behind the lens.

## 5. Constructing lens ray diagrams to predict the superposition of elemental images

With rays connecting elemental images, we can construct ray diagrams to predict where the elemental images compose a sharp image. Like Kepler's ray diagrams [1,2], ours will be based on focal points. Let us re-define focal points in terms of elemental images:
- The front focal point for a convex lens is the place of an object on the optical axis whose elemental images anywhere behind the lens have a separation equal to the pinhole separation; we can find that place during the experiment by varying the position of the object.
- The back focal point for a convex lens (or the front focal point for a concave lens) is the point where each elemental image (from a given viewpoint on the lens) represents an object point along the horizontal line of sight (proceeding from that viewpoint); we can find that point after the experiment by drawing the corresponding rays into the side-view representation, see Fig. 9(e).

Both focal points have the same distance from the lens, which is defined as the so-called focal length [5]. Further, the front and back focal planes are defined as those planes that are one focal length before and behind the lens. Based on these definitions and our quantitative observations from Section 4, we propose the following method for constructing lens ray diagrams:
- *Step 1: Constructing the ray bundle in front of the lens*
  From any viewpoint $P_i$ ($i$ = 1, 2, 3…) on the lens, draw a horizontal ray toward the object, and a ray bundle containing the object, see Fig. 10(a), *cf.* Fig. 11(a).
- *Step 2: Constructing the ray bundle behind the lens*
  From the same viewpoint, draw a focal ray behind the lens. For a convex (concave) lens, the focal ray goes through the back (front) focal point. To obtain the refracted ray bundle, transfer the distances $u_i$ and $l_i$ from the front focal plane into the back focal plane, but in reverse order, see Fig. 10(b), *cf.* Fig. 11(b).
- *Step 3: Locating the complete overlap of ray bundles*
  Do steps 1 and 2 for at least one more viewpoint. Draw the lens image where the refracted ray bundles overlap completely, see Fig. 10(c), *cf.* Fig. 11(c).

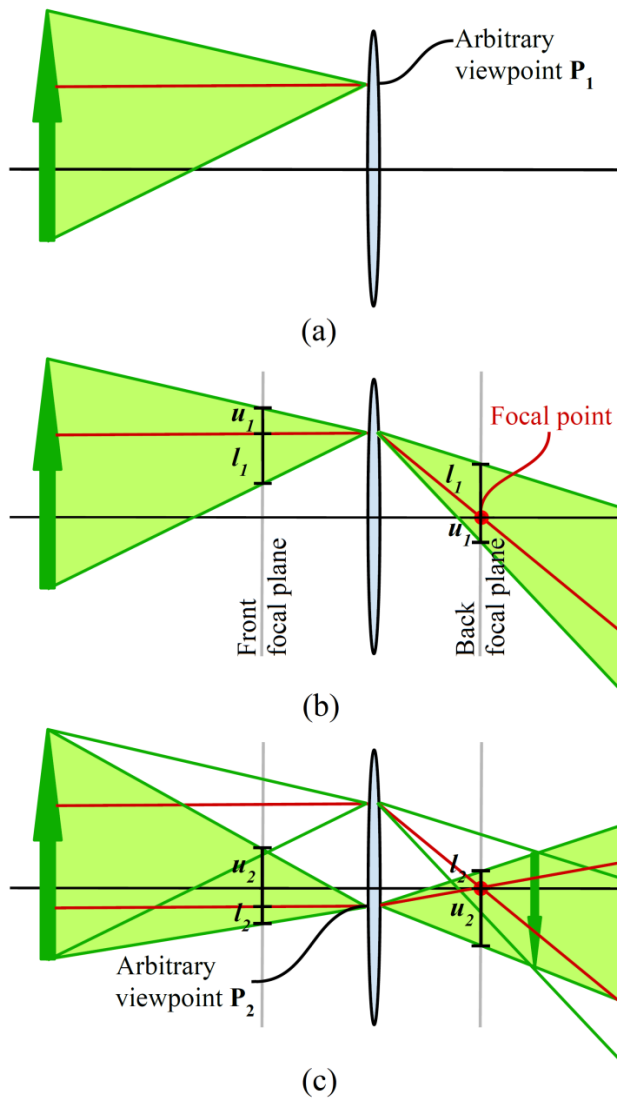

Fig. 10. Constructing a ray diagram for a convex lens, based on ray bundles containing elemental images. (a) From an arbitrary viewpoint $P_1$ on the lens, a horizontal ray and a ray bundle containing the object are drawn. (b) The refracted ray bundle is constructed by transferring the distances $u_1$ and $l_1$ from the front focal plane to the back focal plane, *cf.* Fig. 1(c). (c) The procedure is repeated for another viewpoint $P_2$. The composite image is sharp where the ray bundles from $P_1$ and $P_2$ overlap completely.

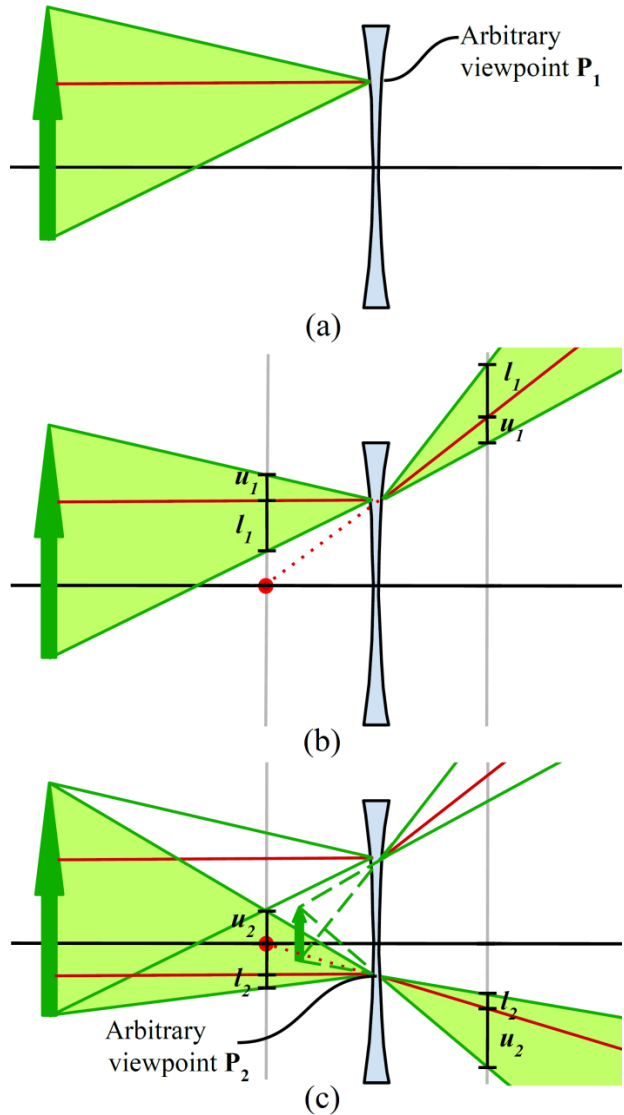

Fig. 11. Constructing a ray diagram for a concave lens, *cf.* Fig. 10. The ray bundles do not completely overlap behind the lens, but if traced backwards, they do in front of the lens, creating a virtual image.

## 6. Discussion

We have treated lens imaging as a superposition of images from different viewpoints. The simulations presented in Section 3 have a digital counterpart in Synthetic Aperture Integral Imaging (SAII) [23,24,28]: In the pick-up stage, the scene is captured with a dense camera array. In the reconstruction stage, the camera images are computationally superimposed by projecting them backwards through a virtual pinhole array. SAII allows computer vision experts to reconstruct a three-dimensional scene from the corresponding image space, and to see through occlusions thanks to the synthesized defocus blur [24]. Using SAII, *Google Inc.* has recently introduced a cell phone app called *Lens Blur*: The user takes a series of photos while moving the camera; afterwards, the app generates the desired defocus blur, called *bokeh* [29].

Although lens imaging and SAII are qualitatively similar, there are notable terminological and quantitative discrepancies: In lens imaging, the term 'focal plane' refers to the plane where the elemental images of infinitely distant objects are completely overlapping; in SAII, it refers to the plane where the elemental images of any object of interest are completely overlapping. This terminological discrepancy reflects the fact that the image space in SAII is congruent with the object space [23,24,28], whereas the image space behind a lens is distorted along the optical axis [14,16]. Consequently, the reconstruction stage in SAII is based on diagrams and formulas that are not applicable to lens imaging.

In Section 4, we have introduced rays as purely geometric constructs, *cf.* [20,30]. As such, they are open to interpretation: In our multi-view approach, rays in front of the lens represent lines of sight [30] proceeding from a point on the lens; consequently, ray bundles in front of the lens represent visual cones, corresponding to many students' preconceptions about light and vision [31]. Likewise, rays behind the lens may be interpreted as lines of sight, or, alternatively, as lines of light going toward an elemental image. In accord with the students' holistic approach [18-21], all rays can be visualised as carrying elemental images from the object to the screen. In accord with Kepler's point-to-point approach [1,2], the rays can be reinterpreted as lines of light.

The re-definitions of focal points in Section 5 are more practical than the conventional definitions: It is always possible to place an object at a finite distance in front of the lens or to identify perfectly horizontal lines of sight (as required by our definition), but it is impossible to have a point source or an infinitely distant object (as required by the conventional definition [1,2]).

The construction method proposed in Section 5 represents refracted camera obscura projections, *cf.* [22]. We have constructed the shift of each camera obscura image based on the deflection of horizontal rays (conventionally called parallel rays), *cf.* [32]. This deflection is known as *prismatic effect* [6], because a lens can be thought of as an array of prisms [6,23], or prism-pinhole pairs [32].

Including the point-to-point construction [12] as a special case, our method has the same limitations as the conventional one. It is only valid for paraxial rays and for a thin lens [5]. Geometric and chromatic aberrations—observable as an imperfect overlap of elemental images—are neglected. Likewise, diffraction is neglected.

Our construction method provides several scientific and pedagogic benefits. First, it is based on concrete phenomena rather than abstract concepts. Second, it implies that any point on the lens can create a complete image, whereas many students think that partially covering the lens would partially destroy the image [18,20]. Third, it includes only those rays that actually pass the lens. Finally, it does not over-emphasize the principal rays, *cf.* [15].

## 7. Conclusion

For the first time since Kepler's approach to lens imaging, we have developed a teaching approach that is based on the different views projected by different points of the lens. We have uncovered these so-called elemental images by covering the lens with a pinhole array. In superposition, these two-dimensional images compose a three-dimensional image. When projected onto a screen, this composite image looks blurry except where the elemental images coincide. Accordingly, we have simulated lens imaging as a superposition of multiple views. Our hands-on simulations have a digital counterpart in Synthetic Aperture Integral Imaging (SAII), allowing teachers to relate the principles of lens imaging to analogous applications in computer vision.

In line with students' preconceptions, we have introduced rays as connections between elemental images. We have proposed a method of constructing lens ray diagrams based on bundles of elemental images. Our method includes the conventional method as a special case.

We have proceeded from the student's own eye to an artificial lens setup, from concrete images to abstract rays, from qualitative descriptions to quantitative predictions, and from a general construction method to the conventional one. Hence, the presented approach may help students to adapt their preconceptions toward the scientific concepts.

**Acknowledgements**

I thank some two dozen 12$^{th}$-grade high school students at the Einstein Gymnasium in Potsdam/Germany for participating in the course "Scharfe Bilder" in February 2015, where we took the proposed multi-view approach. Moreover, I am thankful for constructive remarks by Prof. Dr. Jan-Peter Meyn and Dr. Oliver Passon. I am grateful to Prof. Dr. Florian Theilmann for providing useful feedback throughout my research.